\documentclass[showpacs,twocolumn]{revtex4}
\usepackage[dvips]{graphicx}
\begin{document}
\title{Josephson vortex interaction mediated by cavity modes: \\
Tunable coupling for superconducting qubits}
\author{M. V. Fistul and A. V. Ustinov}
\affiliation{Physikalisches Institut III, Universit\"at
Erlangen-N\"urnberg, D-91058 Erlangen, Germany}

\date{\today}
\begin{abstract}
A quantum-mechanical model for the interaction of Josephson
vortices  (fluxons) embedded in superconducting transmission line
is presented. The vortices interact through emission and
absorption of linear waves (electromagnetic cavity modes). We show
that in a classical regime this peculiar type of interaction is
determined by the product of {\it instantaneous velocities} of
fluxons. In a quantum regime, this property provides tunable
coupling between vortices which can be used for entanglement of
vortex qubits. The physical mechanism of the vortex interaction is
similar to that proposed for qubits based on trapped ions.
Different types of transmission lines mediating the vortex
interaction are proposed.
\end{abstract}


\maketitle

It is well established that Josephson vortices (magnetic fluxons)
can be trapped in spatially extended Josephson structures
\cite{Barone,UstRev}. Moreover, in the presence of an externally
applied magnetic field and dc bias these peculiar nonlinear
objects propagate along a system or are trapped around a
particular point \cite{UstRev,GrEn,UstMalTh,FistUstLibSt}. In a
classical regime the vortex motion is mapped to the classical
mechanics of a macroscopic particle. As we turn to a quantum
regime realized at low temperatures and for a small energy of the
vortex in an external potential, the vortex is predicted to behave
quantum-mechanically, e. g. show tunneling effect and discrete
energy levels \cite{Kato96,Shnirman97b}. Quantum tunneling of
vortices has been observed in discrete arrays of small Josephson
junctions \cite{Zant}, and recently quantum tunneling of a single
vortex was observed in long annular Josephson junctions
\cite{Qtannunp}. A quantum state of a macroscopic vortex in an
artificially created potential can be used in order to implement a
particular type of superconducting qubits named vortex qubits
\cite{WFULowTemp,KempHerSem}.

\begin{figure}
\centering
\includegraphics[width=2.3in]{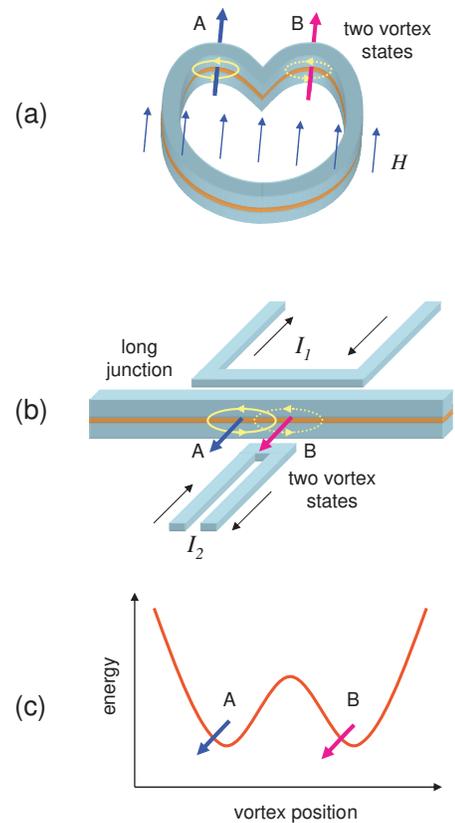}
\vspace{2pt}\caption{Possible realizations of vortex qubits. a)
Heart-shaped long Josephson junction placed in the magnetic field
$H$ \cite{KempHerSem}. b) Long Josephson junction with local
magnetic field injectors. c) Double-well potential for a vortex in
the above two systems. Two vortex states are indicated as A and B.
} \label{fig1}
\end{figure}

Up to now, two types of vortex qubit prototypes have been
investigated. The first prototype is the heart-shaped Josephson
junction shown in Fig.~\ref{fig1}a, for which rather simple state
preparation and readout procedures have been already verified
experimentally \cite{KempHerSem}. In these qubits the external
magnetic field is uniform and the spatially-dependent potential
profile for a vortex is tailored by shaping the long junction into
a heart form. The second vortex qubit idea is based on creating a
desired potential for the vortex by using local
magnetic fields \cite{Ust-APL-2002} generated by control currents
$I_1$ and $I_2$, see Fig.~\ref{fig1}b. Both cases can be
described by a local double-well potential $U(y_i)$ for the vortex,
see Fig.~\ref{fig1}c. Here, $y_i$ is the coordinate of the
center of the vortex.

The qubit shown in Fig.~\ref{fig1}b has the advantage that its
state preparation and readout can be done by using Rapid Single
Flux Quantum (RSFQ) circuits. RSFQ logic is the natural choice for
an interface between vortex qubits and room-temperature
electronics. The field-injector based vortex qubits with  RSFQ
preparation and read-out interfaces have been also recently tested
in the classical regime \cite{Ust-Kapl-2002}. The entanglement
that persists in the quantum regime in a multi-qubit circuit
should be possible to read out by RSFQ circuitry. The vortex
qubits are naturally scalable to larger quantum circuits.

\begin{figure}
\centering
\includegraphics[width=3.5in]{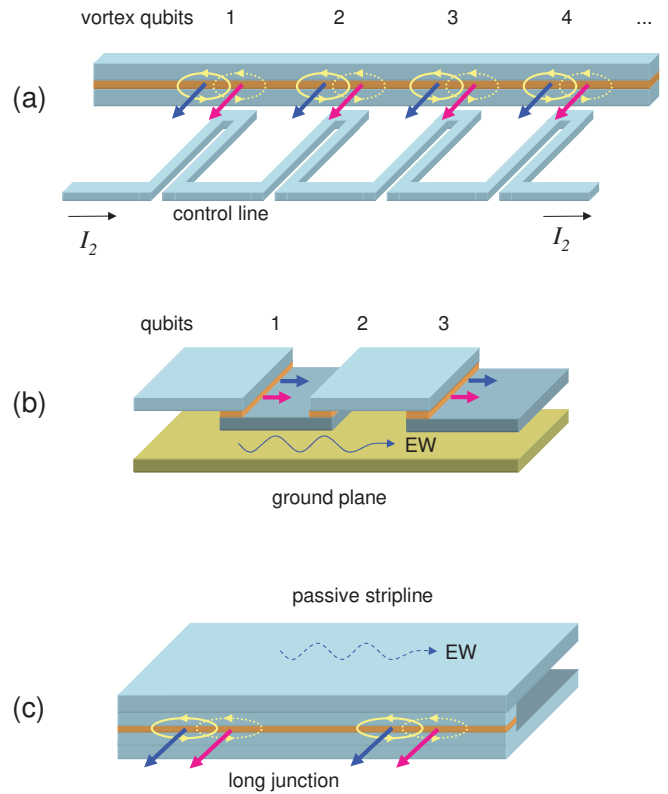}
\vspace{2pt}\caption{Sketch of long Josephson junction circuits
allowing for different types of vortex interaction. a) Static
coupling between vortex qubits can be realized by placing them
side by side in a long junction. The distance d between qubits has
to be larger (but not much larger) than $\lambda_J$. b) The
dynamic coupling between vortex qubits located in different
junctions. The junctions are placed above a superconducting ground
plane. c) The dynamic coupling between vortex qubits separated by
large distance in a single long junction. Note a passive stripline
behind the junction. It provides the coupling between qubits via
electromagnetic waves (EW) propagating in the stripline. }
\label{fig2}
\end{figure}

In a system containing a number of vortices they interact with
each other. The simplest type of the interaction, namely {\it
direct static} magnetic interaction, occurs when the vortices are
located in the same long Josephson junction, see Fig.~\ref{fig2}a.
In this case the interaction is due to the presence of a mutual
inductance between the two vortex locations and is determined by
overlapping of the "tails" of vortex magnetic fields. Such an
interaction decays exponentially \cite{comment} with the distance
between the vortices on the characteristic scale of the Josephson
penetration depth $\lambda_J$ and depends on their {\it
coordinates}. In the quantum regime the coupling between vortices
leads to a quantum entanglement, which is the necessary condition
in order to realize a quantum computation.

The static interaction exponentially depends on the distance
between vortices so that only coupling between the nearest
neighbors has to be taken into account. In coordinate
representation the static interaction term between two vortices
can be written as $U_0\exp({|y_i-y_j|/\lambda_J})$, where $y_i$,
$y_j$ are the coordinates of the interacting vortices, and $U_0$
is the characteristic interaction energy. If we switch to the
representation of wave functions corresponding to the two energy
levels of single vortices, the interaction term in the Hamiltonian
can be written as $\hat{H}_{\rm int}=\sum
U_0\exp({|a_i-a_j|})\hat{\sigma}^i_z\hat{\sigma}^j_z$, where
$a_i$, $a_j$ are the \textsl{average} coordinates of corresponding
vortices normalized to $\lambda_J$.  Note here, that with this
type of interaction the chain of vortices is mapped to a 1-D Ising
model of interacting spins $1/2$ or to a system of nuclear spins
\cite{Lloyd}.

A severe drawback of the static type of interaction is that it can
not be turned off and addressing of a chosen qubit (pinned vortex)
will immediately affect all other qubits. Thus, in this
realization it is easy to obtain entangled states but it remains a
hard problem to address a single vortex, e.g. switch from the
ground state to the excited state ($|0> \rightarrow |1> $
transition) of a particular vortex without affecting others.

In order to avoid the above problem, we propose for coupling of
vortex qubits placing them in a superconducting transmission line
(Fig.~\ref{fig2}b and \ref{fig2}c). In this case the vortices
display a peculiar { \it indirect dynamic} interaction through
virtual excitation (absorption) of linear electromagnetic waves
(EW) propagating in a transmission line. This interaction can be
realized at least in two different configurations. One is a
lateral configuration shown in Fig.~\ref{fig2}b, where two or more
long Josephson junctions are incorporated in a common system with
a superconducting ground plane. In such a system the
superconducting electrodes and the ground plane provide a low
dissipative transmission line for propagation of EW. Thus, a {\it
moving} (oscillating) vortex excites the EW (emission of photons)
which propagate through the transmission line and interact (by
absorption of photons) with other vortices. Note, that this
principle allows to realize a quantum-mechanical interaction
similar to the one proposed for chains of trapped ions
\cite{BDMTintr,Iontrap}. In the latter system single-qubit
operations address ion energy levels but the interaction is just
due to the excitation of phonons in a chain of trapped ions.
Similar realization of coupling via lumped $L$-$C$ circuit has
been proposed earlier \cite{Makh,Makhreview} for superconducting
qubits.

Anther possible realization of the dynamic coupling is a
configuration with far-separated vortices located in the same
Josephson junction. These vortex qubits indirectly interact with
each other through the EW excitations in a passive transmission
line inductively coupled to the junction, see Fig.~\ref{fig2}c.
Similarly to a lateral configuration, the EW propagating in
transmission line provides a fast and low dissipative source of
interaction.

In the following, we  quantitatively analyze the lateral
configuration of vortex qubits located in different long junctions
placed above a superconducting ground plane (Fig.~\ref{fig2}b). A
{\it classical dynamics} of such a system is determined by
time-dependent Josephson phases $\varphi (y-y_{i},t)$, where $y$
is the coordinate along Josephson junctions and $y_i$ is the
coordinate of the center of the vortex located in the $i$-th
Josephson junction. The EW propagation along the transverse
coordinate $x$ is described by a set of equations:
$$
\frac{\partial V(x,t)}{\partial x}~=~L\frac{\partial
I(x,t)}{\partial t} +R_sI(x,t) +
$$
$$
+\sum_i \int \frac{dy}{a}
\frac{\hbar}{2e}\frac{\partial \varphi(y-y_i,t)}{\partial t} \delta(x-x_i)
$$
\begin{equation} \label{EquationEWs-0}
 \frac{\partial I(x,t)}{\partial x}~=~C\frac{\partial V(x,t)}{\partial t}~~,
\end{equation}
where $x_i$ are the coordinates of Josephson junctions, $a$ is the
width of the Josephson junction in $x$- direction, $V(x,t)$ is the
voltage between the superconducting electrodes and the ground
plane, $I(x,t)$ is the current flowing in $x$ direction along the
superconducting electrodes. The parameters $L$ (the inductance of
superconducting electrodes per unit length) and $C$ (the
capacitance between superconducting electrodes and the ground
plane per unit length) determine the properties of a transmission
line created between the superconducting electrodes and the ground
plane. The parameter $R_s$ describes the dissipation due to the
surface losses (in general, they are frequency dependent) in the
transmission line. For not very high frequencies these losses in
the superconducting line are very low. Notice here, that without
the term depending on the Josephson phases this set of equation
describes the propagation of a weakly decaying linear EW with the
spectrum $\omega(k)~=~ck$, where $c~=~1/\sqrt{LC} $
\cite{Feinman}. In the presence of vortices we obtain the wave
equation where the vortices are the local sources of perturbation:
$$
\frac{\partial^2 I(x,t)}{\partial x^2}-CL\frac{\partial^2 I(x,t)}
{\partial t^2} - RC\frac{\partial I(x,t)}{\partial t}~=~
$$
\begin{equation} \label{EqCurrdistr}
 ~=~-C\frac{\pi\hbar}{ae}
 \sum_i  \ddot y_i \delta(x-x_i)~~.
\end{equation}
Thus, we obtain a time Fourier transformation of the ac current
distribution in the transmission line
$I_\omega (x)$ in the form:
\begin{equation} \label{EquationEWs}
 I_\omega (x)~=~C\frac{\pi\hbar}{ae}\sum_i \omega^2 G_\omega
 (x,x_i) \int dt e^{i\omega t}y_i(t)~~,
\end{equation}
where $G_\omega(x,x_i)\equiv G_\omega(x-x_i)$ is the Green
function of the linear equation:
\begin{equation} \label{Grfunct}
 \frac{d^2 G_\omega(x,x_1)}{d x^2}+CL\omega^2G_\omega(x,x_1)
 -i\omega RC G_\omega(x,x_1)=
 \delta (x-x_1).
\end{equation}

The equation of motion for a vortex center of mass in the presence
of a local double-well potential $U(y_i)$ illustrated in
Fig.~\ref{fig1}c is written as
\begin{equation} \label{FlEqMot}  \ddot y_i +\frac{\pi \lambda_J
\omega_p^2}{4}\left[\frac{2e\lambda_J}{\hbar I_{c0}}
\frac{\partial U(y_i)}{\partial y_i}-
\frac{I(x_i,t)}{I_{c0}}\right]~=~0~~,
\end{equation}
where $\omega_p$ and $I_{c0}$ are the plasma frequency and the
critical current of a Josephson junction (in the absence of
vortices), respectively \cite{GrEn,UstMalTh}. Thus, the
interaction energy between vortices can be found as:
\begin{equation} \label{Interaction}
 U_{int}~=~-\sum_i \frac{\pi\hbar}{e} \frac{I(x_i,t) y_i(t)}{\lambda_J}~~.
\end{equation}
By making use of (\ref{EquationEWs}) and (\ref{Grfunct})
we obtain
$$
U_{int}~=~ \frac{C}{a\lambda_J}\left(\frac{\pi \hbar}{e}\right)^2
$$
\begin{equation} \label{Interaction2}
 \times \sum_{i \neq j} \int dt_1 \dot y_i(t) \dot y_j(t_1)\int
 d\omega e^{{\rm i}\omega (t-t_1)}
 G_\omega (x_i-x_j) ~~.
\end{equation}
This expression can be symmetrized over the indexes $i$ and $j$ as
$$
U_{int}~=~ \frac{2\tau C}{a\lambda_J}\left(\frac{\pi
\hbar}{e}\right)^2
$$
\begin{equation} \label{Interaction3}
  \times \frac{1}{2}\sum_{i \neq j} \int dt_1 \dot y_i(t) \dot y_j(t_1)
  K(t-t_1, |x_i-x_j|)~,
\end{equation}
where the kernel of interaction
\begin{equation} \label{Kernel}
 K(t,x)~=~-{\rm i} \int d\omega \omega e^{-{\rm i}\omega t}
 G_\omega (|x|) ~~
\end{equation}
depends on the parameters $L$ and $C$ in our model. In the limit
of a small propagation time ($\tau ~=~ |x_i-x_j|\sqrt{LC} \ll
T_{\rm osc} $, where $T_{\rm osc}$ period of slow vortex
oscillations between two quantum states) we neglect a delay in the
wave propagation, and obtain the {\it classical} interacting
potential between vortices located in the Josephson junctions $i$
and $j$:
\begin{equation} \label{QLInteractionFin}
 U_{ij}~=~A_{ij}\dot y_i(t) \dot y_j(t)~~,
\end{equation}
where the constant of interaction $A_{ij}$ is determined by the
properties of the transmission line given by (\ref{Kernel}). For a
simplest case of a transmission line described by
(\ref{EqCurrdistr}) we obtain
$$
A_{ij}~=~\frac{|x_i-x_j| C}{a\lambda_J}\left(\frac{\pi
\hbar}{e}\right)^2
\exp\left({-\frac{R}{2}\sqrt{\frac{C}{L}}|x_i-x_j|}\right)~.
$$
Thus, in the case of an indirect dynamic coupling the vortex
interaction is determined by a product of {\it vortex velocities}.

In the {\it quantum regime} this dynamic interaction is determined
by operators of vortex {\it momentum} ($\hat p_i$ and $ \hat
p_j$):
\begin{equation} \label{InteractionQuantum-2}
 \hat H_{int} ~=~\frac{ A_{ij}}{m_f^2}\hat p_i \hat p_j~~,
\end{equation}
where $m_f$ is the effective mass of a vortex \cite{WFULowTemp}.
The total Hamiltonian $H$ of indirectly interacting vortices can
be written as
\begin{equation} \label{InteractionQuantum}
 \hat H ~=~\sum_i \frac{\hat p_i^2}{2m_f}+ U(y_i) +\frac{1}{2}\sum_{i \neq j}
 \frac{ A_{ij}}{m_f^2}\hat p_i \hat p_j~~.
\end{equation}

If a single vortex potential $U(y_i)$ has a double-well form, the
quantum-mechanical description of a single vortex can be reduced
to lowest energy levels in each well. These levels are
characterized  by the energy difference  $\sim B_z$ and the
tunneling amplitude $\sim B_x$ between states \cite{WFULowTemp}.
The parameters $B_z$ and $B_x$ are determined by the properties of
a single vortex potential $U(y_i)$. Because the average value of
the momentum operator $<\hat p_i>$ is zero in these states, we
obtain that the interaction potential is determined by tunneling
amplitudes $B_x(i)$. By making use of the spin representation of
the Hamiltonian we write
\begin{equation} \label{InteractionQuantumSpin}
 \hat H ~=~\sum_i \left [ B_z(i) \hat \sigma^i_z +B_x(i) \hat \sigma^i_x \right ] +
  \frac{1}{2}\sum_{i \neq j} B_x(i) B_x(j) \tilde A_{ij} \hat \sigma^i_x \hat \sigma^j_x~.
\end{equation}
A great advantage of this dynamic type of interaction in respect
to a direct static one is that the former one is absent when both
vortices are in the ground state. With this type of Hamiltonian it
is possible to address a single vortex or a bunch of vortices, if
that is required \cite{Makhreview}. Indeed, by turning off the
tunneling amplitude $B_x$ of a particular vortex we exclude it
from the computation process, but other vortices can still be
manipulated.

In conclusion, we proposed here and quantitatively analyzed an
interaction of vortices through an exchange of EW propagating in a
superconducting transmission line. In a classical regime, this
interaction depends on the velocities of vortices. As we turn to a
quantum regime, this type of interacting allows for a
\emph{tunable coupling} between vortices that is a crucial
condition for implementation of such vortices as superconducting
qubits. The addressing of these qubits can be provided by a particular
combination of dc pulses allowing to change the parameters $B_z$ and $B_x$.

We thank A. Kemp for useful remarks on the manuscript and
acknowledge discussions with V. K. Kaplunenko, J. Lisenfeld, and
A. Wallraff.


\begin{references}
\bibitem{comment} The interacting vortices  can be also located in
different long Josephson junctions. In this case the direct static
magnetic interaction is characterized by the power law but the
length of the interaction is much smaller (of the order of the
London penetration depth $\lambda_L$, as the thickness of the
electrodes $d~\simeq~\lambda_L$).
\end{references}
\end{document}